\documentclass[journal=jacsat,manuscript=article]{achemso}

\usepackage[version=3]{mhchem} 
\usepackage[ruled,vlined]{algorithm2e}
\usepackage{amsmath}
\usepackage{float}
\usepackage{subfigure}
\usepackage{natbib}
\usepackage{graphbox} 
\usepackage{mwe}
\usepackage{graphics}
\usepackage{graphicx}
\usepackage[export]{adjustbox}
\usepackage{pgfplots}

\newcommand{\shiftleft}[2]{\makebox[0pt][r]{\makebox[#1][l]{#2}}}
\newcommand{\shiftright}[2]{\makebox[#1][r]{\makebox[0pt][l]{#2}}}

\pgfplotsset{compat=1.14}

\author{Maxime Langevin}
\affiliation{PASTEUR, D\'epartement de chimie, Ecole Normale Supérieure, PSL University, Sorbonne Université, CNRS, 75005 Paris, France}
\alsoaffiliation[Sanofi]
{Molecular Design Sciences - Integrated Drug Discovery, Sanofi R\&D, Vitry-sur-Seine, France}
\author{Herv\'e Minoux}
\affiliation[Sanofi]
{Molecular Design Sciences - Integrated Drug Discovery, Sanofi R\&D, Vitry-sur-Seine, France}
\author{Maximilien Levesque}
\affiliation{PASTEUR, D\'epartement de chimie, Ecole Normale Supérieure, PSL University, Sorbonne Universit\'e, CNRS, 75005 Paris, France}
\alsoaffiliation{Aqemia, Paris, France}
\email{maximilien.levesque@aqemia.com}
\author{Marc Bianciotto}
\affiliation[Sanofi]
{Molecular Design Sciences - Integrated Drug Discovery, Sanofi R\&D, Vitry-sur-Seine, France}
\email{marc.bianciotto@sanofi.com}

\title[]
  {Scaffold-constrained molecular generation}

\abbreviations{IR,NMR,UV}

\begin{document}

\section{Abstract}

One of the major applications of generative models for drug discovery targets the lead-optimization phase. During the optimization of a lead series, it is common to have scaffold constraints imposed on the structure of the molecules designed. Without enforcing such constraints, the probability of generating molecules with the required scaffold is extremely low and hinders the practicality of generative models for \textit{de-novo} drug design. 
To tackle this issue, we introduce a new algorithm to perform scaffold-constrained \textit{in-silico} molecular design. We build on the well-known SMILES-based Recurrent Neural Network (RNN) generative model, with a modified sampling procedure to achieve scaffold-constrained generation. 
We directly benefit from  the associated reinforcement learning methods, allowing to design molecules optimized for different properties while exploring only the relevant chemical space.
We showcase the method’s ability to perform scaffold-constrained generation on various tasks: designing novel molecules around scaffolds extracted from SureChEMBL chemical series,  generating novel active molecules on the Dopamine Receptor D2 (DRD2) target, and finally, designing predicted actives on the MMP-12 series, an industrial lead-optimization project.

\section{Introduction}
Finding new drugs is a long, costly and difficult \cite{Schneider2016} problem, with potential failure all along the drug discovery pipeline and an overall success rate \cite{Paul2010} close to only 4\%. Lead-optimization, where medicinal chemists refine bioactive molecules into potential drugs, takes a major part of the time and cost in the discovery pipeline. Finding a good drug candidate requires finding a molecule that is active on the target of interest while satisfying multiple criteria mainly related to safety and pharmacokinetics. In this respect, lead-optimization can be viewed as a multi-objective optimization problem in chemical space.
There has been a recent surge of interest in generative models for \textit{de-novo} drug design \cite{C9ME00039A} and their application in the drug discovery pipeline. Generative models have been studied for two types of tasks: distribution learning and goal-oriented learning \cite{Brown2019}. 
Distribution learning aims at reproducing the distribution of a known dataset, in order to sample a large library of molecules similar to the initial dataset used to train the generative model.
Goal-oriented learning, on the other hand, takes as input a scoring function and aims at finding the molecules with the highest scores. Applying generative models to lead-optimization can actually be understood as a special case of goal-oriented learning, where the scoring function reflects the adequacy of the molecule to the different project objectives. Distribution learning benchmarks are also frequently used to assess whether a model has learnt to generate drug-like molecules, and will be a good starting point for goal-oriented learning.

However, real life lead optimization projects very often impose constraints on the scaffold of designed molecules. Interesting scaffolds are identified during the lead identification phase of the pipeline, and often kept throughout the rest of the pipeline, with only minor changes. The scaffold, or “core” of the molecule, is key to (i) preserving biological activity identified earlier in the pipeline (ii) staying within areas of chemical space where prior information gathered is relevant, making SAR (structure-activity relationship) understandable (iii) maintaining a high throughput for compound synthesis by using common precursors, speeding the Design-Make-Test-Analysis (DMTA) cycle to a level acceptable by industry standards and (iv) exploring a relevant chemical space, translatable into a Markush structure, from an Intellectual Property perspective.

Therefore, lead-optimization is actually most of the time \cite{Hughes2011} a multi-objective optimization problem under scaffold constraints. Yet, goal-oriented learning with scaffold constraints is not extensively studied, while being of major interest in the application of generative models to medicinal chemistry.
As showed in our work, simply including the presence of the scaffold of interest in the scoring function as a supplementary criterion and applying existing goal-oriented models \cite{Sthl2019} shows poor results. Hence, there exists a clear need for algorithms that include directly those constraints in their generative process.

Before the increasing rise of interest \cite{C9ME00039A} in generative models for chemistry, traditional approaches were already studied by computational chemists for \textit{de-novo} design of optimized molecules. Those approaches relied either on combination of fragments and exploration of chemical space by performing virtual reactions with handcrafted rules \cite{doi:10.1002/wcms.49}. Genetic algorithms, that showed success in many optimization tasks, were also adapted to chemistry and used in order to design molecules of interest \cite{genetic, C8SC05372C}.
Following  Gomez-Bombarelli et al.\cite{Gomez_Bombarelli_2018} who applied recent generative models in order to generate new molecules, there has been increasingly growing interest \cite{C9ME00039A, Schneider2019} for \textit{de-novo} design with generative models. Indeed, relying on a data-driven method that learns the underlying distribution of molecules helps to remain in a chemical space of drug-like molecules. 

Among the various molecular generative models published in recent years, many actually rely \cite{C9ME00039A} on RNN, and especially on the Long Short Term Memory (LSTM) architecture. RNN  are frequently used as generative models for sequential data, and in particular for SMILES \cite{doi:10.1021/acscentsci.7b00512, article}. Those models differ by the way they aim at learning the probability distribution of the training set. Variational Auto-Encoders (VAE) \cite{Gomez_Bombarelli_2018,DBLP:journals/corr/abs-1811-12823} learn this distribution through an encoder-decoder loss, where the generator is used as a decoder and another RNN used as encoder. Generative Adversarial Networks (GAN) \cite{sanchez-lengeling_outeiral_guimaraes_aspuru-guzik_2017,DBLP:journals/corr/abs-1811-12823}  learn the distribution through a competition between the generator and a discriminator, and character-level generator \cite{doi:10.1021/acscentsci.7b00512,article} through simple negative log-likelihood minimization. While the methods to learn the distribution differ, in the end those methods rely on the same architecture to generate new molecules, which is SMILES-based RNN.
To design optimized molecules with RNN, the easiest way is to generate a large virtual library of molecules and then screen it against the objectives, but the likelihood of finding optimized molecules is extremely small.
Reinforcement Learning (RL) was proposed as a way to design optimized molecules \cite{doi:10.1021/acscentsci.7b00512}. Either through policy gradient or hill-climbing, which both aim at maximizing the probability of the highest scoring SMILES, the RNN can be fine-tuned to generate high scoring molecules. Hill-climbing was shown \cite{Brown2019} to achieve state-of-the-art results on various goal-oriented benchmarks.
Optimizing directly on the molecular graph, controlled either by a neural network \cite{Zhou2019} or through Monte-Carlo Tree Search \cite{C8SC05372C}, can yield interesting results \cite{Brown2019}, but also leads to compounds of lesser \cite{Brown2019} quality, which hinders adoption of those approaches for application in drug discovery projects.
Methods based on Graph Neural Networks, that work directly on the molecular graph rather than with SMILES \cite{pmlr-v80-jin18a,DBLP:journals/corr/abs-1805-11973}, have also been studied recently. Those architectures are more experimental and haven’t proved for the moment to perform as well as SMILES based methods.

Scaffold-constrained generation has been investigated in only a few previous works. Fragment growing with RNN \cite{Merk2018, Gupta2017} has been described, but is very limited in its scope and cannot accommodate scaffold constraints besides simply imposing the presence of a fragment. DeepScaffold \cite{Li2019} relies on Graph Neural Networks to build molecules on an existing scaffold. The architecture used is complex and experimental \cite{ArsPous2020}, relying on graph convolutional neural networks. The absence of built-in reinforcement learning also hinders the adoption of this method to design optimized molecules. The Reinvent Scaffold Decorator \cite{ArsPous2020} is based on SMILES and RNN and uses an encoder-decoder architecture. It completes scaffolds in two different ways, single-step and multi-step. Single-step completes all open positions at once, while multi-step completes each open position iteratively. The architecture used is also complex, as it relies on an encoder-decoder scheme, and is limited to open positions at branching points. The model also requires dataset-specific preprocessing and training for each application. Moreover, no RL procedure is directly applicable, limiting as well the scope of this method. 

Prior work on Markov models and especially RNN \cite{pmlr-v77-walder17a, 10.5555/3102787.3102812} showed that for a given language, constraints that can be translated sequentially can be enforced during  sampling. Thus, the main idea behind our model is that scaffold constraints can actually be translated to the SMILES language. Acknowledging this fact, we investigate how we can leverage this translation from structure to SMILES to perform scaffold-constrained generation. We show that scaffold-constrained generation can be achieved with SMILES-based RNN using simply a modified sampling procedure. This means that our method tackles scaffold-constrained generation without the need for a new model, or even for retraining this model. Moreover, all previous experience acquired by computational chemists with SMILES-based RNN usage translates with our procedure.
The available state-of-the-art reinforcement learning methods for goal-oriented learning ensures that our procedure is easily applicable for lead-optimization.
By performing scaffold constrained generation, we can include the project's structural constraints directly in our generative process. This allows us to search directly within the very sparse subspace of chemical space that is relevant to the lead-optimization project. Furthermore, our framework allows low-level control on scaffold-constrained molecular generation, including the possibility to limit oneself to linear fragments, to control the presence of branches and cycles, or the type of atoms authorized, as well as the ability to perform scaffold-hopping. 

We begin by giving  background on molecule generation with SMILES-based RNN. Then, we show how to adapt the sampling procedure in order to perform scaffold constrained generation with an RNN. Finally, we provide  experimental results on different \textit{de-novo} design tasks, and show the usefulness of our method in the context of scaffold constrained generation.

\section{Methods}
The SMILES \cite{smiles} language represents molecules as a chain of characters. Each character corresponds to an atom or denotes structural information, such as opening or closing of cycles and branches, stereochemistry or multiple bonds.

Starting from a molecular graph, cycles are broken down and marked with numbers, and a starting atom is chosen. The SMILES string is obtained by printing the symbol of the nodes encountered in a depth-first traversal starting at the chosen atom, with parentheses indicating branching points in the graph. 
For a given molecular graph, there are as least as many SMILES as possible starting atoms. A canonicalization algorithm \cite{smiles} can be used to pick the starting atom, thus yielding the canonical SMILES of the molecule. The corresponding molecular graph can be easily retrieved from a given SMILES.

RNN can generate SMILES in a sequential fashion by modeling the conditional probability distribution over SMILES tokens (conditioned on the beginning of a SMILES). The SMILES language is enriched with a GO and EOS token \cite{doi:10.1021/acscentsci.7b00512} that denote the beginning and the end of a SMILES string. Let $s=x_{0},....x_{n}$ the tokenized version of a SMILES string, with $x_{i}$ characters from the SMILES language (here $x_{0}$ and $x_{t}$ denote respectively the GO and EOS tokens). The RNN models $P(x_{t}\mid x_{0},... x_{t-1})$, i.e. the conditional probability of a token given all the previous ones. 
The RNN is trained on a database of drug-like molecules (such as ChEMBL \cite{chemBL}) to predict the next token of a SMILES given the beginning of the sequence, and training is achieved through minimization of the negative log-likelihood of the training set SMILES. The objective is to minimize (w.r.t the RNN parameters denoted by $\theta$) the following loss:
\begin{equation}
  L(\theta) = -\sum_{t=1}^T {\log P{(x_{t}\mid x_{0},... x_{t-1})}}   
\end{equation}
The RNN relies on its internal state $h$ to process the information from the previous tokens, and models the conditional probability $P(x_{t}\mid x_{0},... x_{t-1})$ as $P(x_{t}\mid h_{t-1})$. 
Data augmentation by enriching the dataset with non-canonical SMILES \cite{Arus-Pous2019} has shown benefits in different tasks, therefore we adopt it in our methodology when training RNN for the following experiments.

Once the conditional probability distribution $P(x_{t}\mid x_{0},... x_{t-1})$ is learnt, sampling is achieved by initializing the sequence with a GO token, and then sampling tokens sequentially. This procedure ends when a EOS token is sampled, and returns a SMILES with its corresponding likelihood, which is amenable to backpropagation.

\begin{algorithm}[H]
\SetAlgoLined
\KwResult{SMILES string}
 initialize $h_{0}$ \;
 $x_{0}$ = $GO$ \;
 $t = 1$ \;
 \While{$x_{t} \neq EOS$}{
  Sample $x_{t}$ from $P(x_{t}\mid x_{0},... x_{t-1})$ and update $h_{t-1}$ to $h_{t}$\;
  $t = t+1$ \;
 }
 \KwOut{$x_{0},... x_{t}$}
 \caption{Generating new SMILES samples}
\end{algorithm}

This framework can be adapted to scaffold constrained generation. Indeed, a scaffold can be viewed as an incomplete SMILES. This incomplete SMILES can have different positions with open possibilities, denoted by the special token “*”.
An example of how scaffold constraints can be translated in SMILES, in the case of an open branch, is given Table \ref{table: SMILES syntax and molecular structure branch}.

\begin{table}[H]
  \begin{center}
  \begin{tabular}
      {|p{7cm}|p{7cm}|} 
      \hline
      \begin{center}\textbf{Original structure}\end{center} & \begin{center}\textbf{Structure with open position}\end{center} \\
      \hline 
      \hline \shiftright{0pt}{\raisebox{-0.9\height}{
      \includegraphics[scale=0.5]{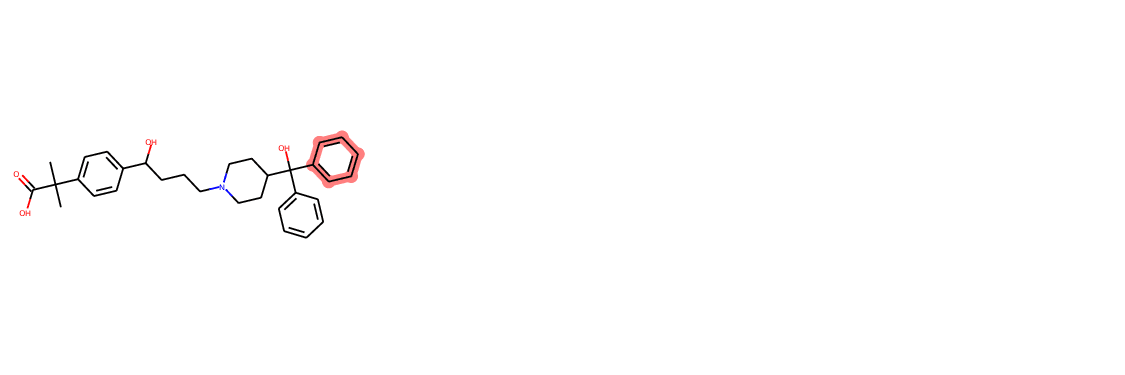}}} &  \shiftright{1pt}{\raisebox{-0.9\height}{
      \includegraphics[scale=0.12]{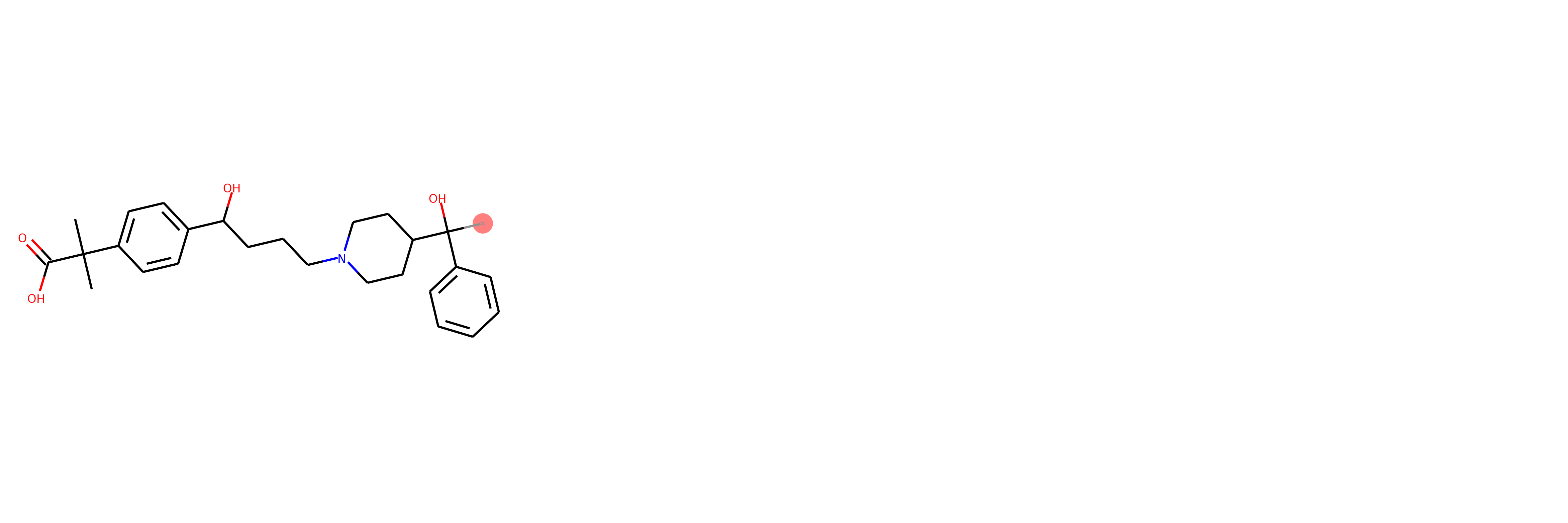}}} \\
      \hline
      \begin{center}CC(C)(C(=O)O)c1ccc(cc1)C(O)CCC\\N2CCC(CC2)C(O)\textbf{(c3ccccc3)}c4ccccc4\end{center} & \begin{center}CC(C)(C(=O)O)c1ccc(cc1)C(O)CC\\CN2CCC(CC2)C(O)\textbf{(*)}c4ccccc4\end{center} \\
      \hline
     
  \end{tabular}
   \caption{Linking SMILES syntax with molecular structure: open branch}
   \label{table: SMILES syntax and molecular structure branch}
   \end{center}
\end{table}

When the open position is within a linker, an example is given Table \ref{table: SMILES syntax and molecular structure linker}.
\begin{table}[H]
  \begin{center}
  \begin{tabular}
      {|p{7cm}|p{7cm}|} 
      \hline 
        \begin{center}\textbf{Original structure}\end{center} & \begin{center}\textbf{Structure with open position}\end{center} \\
      \hline 
      \hline\shiftright{0pt}{\raisebox{-0.9\height}{
      \includegraphics[scale=0.5]{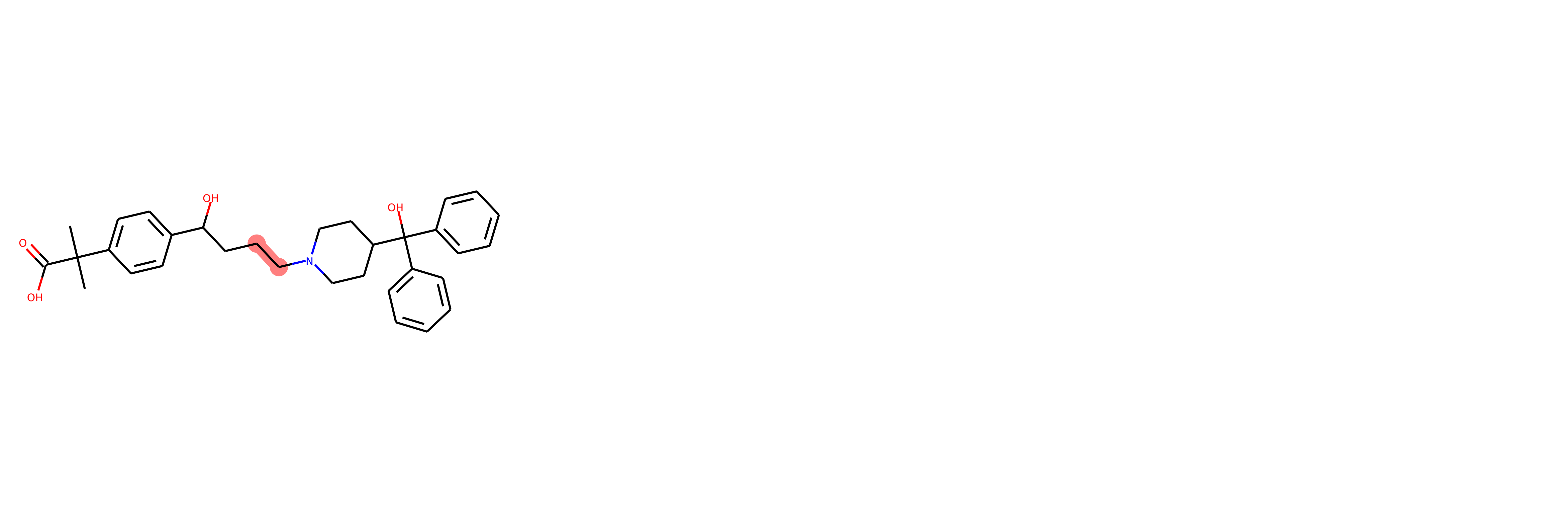}}} &  \shiftright{1pt}{\raisebox{-0.9\height}{
      \includegraphics[scale=0.5]{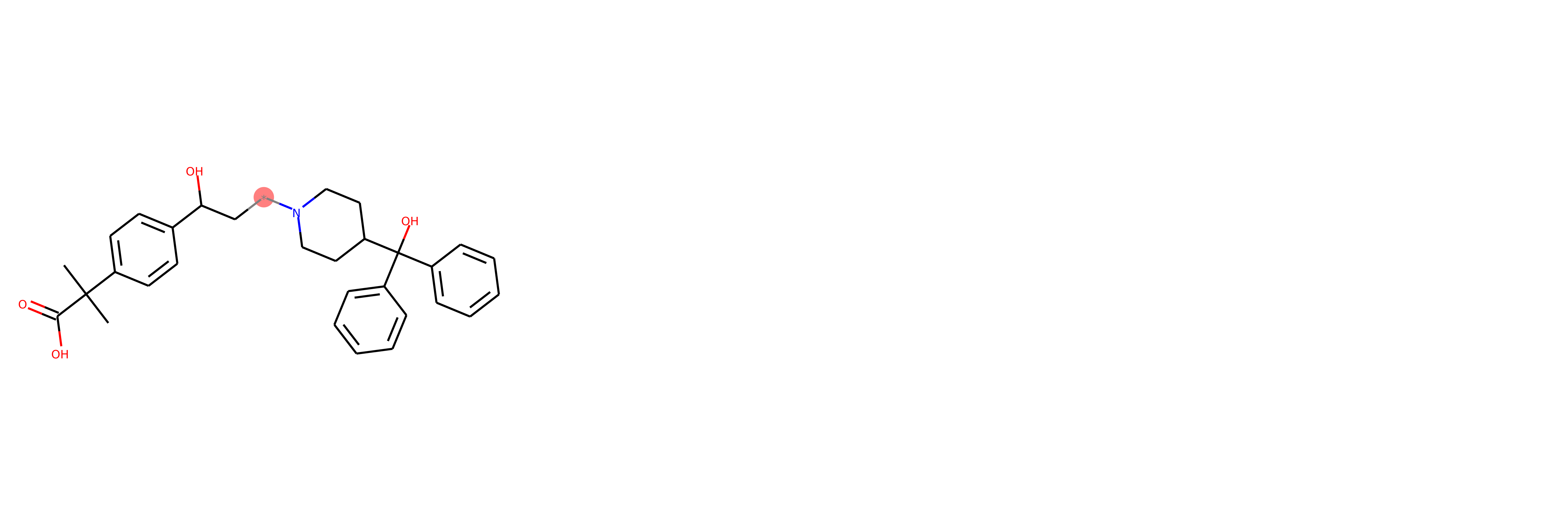}}} \\
      \hline 
      \begin{center}CC(C)(C(=O)O)c1ccc(cc1)C(O)C\textbf{CC}N\\2CCC(CC2)C(O)(c3ccccc3)c4ccccc4\end{center} & \begin{center}CC(C)(C(=O)O)c1ccc(cc1)C(O)C\textbf{*}N2\\CCC(CC2)C(O)(c3ccccc3)c4ccccc4\end{center} \\
      \hline
     
  \end{tabular}
   \caption{Linking SMILES syntax with molecular structure: open linker}
   \label{table: SMILES syntax and molecular structure linker}
   \end{center}
\end{table}
In the classic procedure, the RNN samples freely tokens until the end of the SMILES is reached. For scaffold constrained generation, the RNN is constrained to follow the SMILES scaffold, and sampling is enabled only when an open position “*” of the SMILES is reached. 

The major subtlety lies in the fact that the sampling procedure depends upon the configuration of the molecular graph around the open position, especially with regard to the decision to stop sampling and resume reading. In this work, we tackle three different kinds of open positions. First, open positions at branching points of the molecular graph (commonly referred to as R-groups). Then open positions in linkers (that link different cycles of the molecule), and finally constrained choices, where the position is open but the number of possibilities is already limited within the drug discovery project. Each kind of open position requires an adequate sampling procedure. The general algorithm for scaffold constrained sampling is the following:

\begin{algorithm}[H]
\SetAlgoLined
\KwResult{SMILES string with scaffold $s$}
 \KwIn{scaffold $s=s_{1},...,s_{n}$}
 initialize $h_{0}$ \;
 $x_{0}$ = $GO$ \; 
 $t = 1$ \;
 \For{$i\gets1$ \KwTo $n$}{
  \uIf{$s_{i}$ not $*$}{
   Read $s_{i}$ and update $h_{t-1}$ to $h_{t}$ \;
   $x_{t} = s_{i}$ \;
   $t = t+1$ \;
  }
  \uElse{
   Sample $y = (y_{1},...,y_{k})$ and update $h$ according to special sampling procedure for scaffold decoration\;
   $x_{t},...,x_{t+k} = y_{1},...,y_{k}$ \;
   $t = t+k$
   
  }
  }
  $t = t+1$\;
  $x_{t}$ = $EOS$ \;
  
  \KwOut{$x_{0},... x_{t}$}

 \caption{Generating new SMILES samples with scaffold constraints}
\end{algorithm}

Decorating a fixed scaffold is one the predominant techniques used by medicinal chemists in lead optimization \cite{Bhm2004} and being able to perform this task with a generative model is of major practical importance in their application to drug discovery.
For branched decorations (i.e R-groups), the SMILES translation is straightforward: opening and closing parentheses denote the beginning and end of the branch, and therefore an open branched decoration will translate to “(*)” in the SMILES language. 
In this case, the sampling procedure is easily designed: the RNN is free to sample any tokens while the branch is open. When a closing parenthesis that matches the opening parenthesis of the branch is sampled, then it means the RNN finished to sample the branched decoration, and can resume reading the rest of the scaffold. This also implies that it is necessary to keep track of the opening and closing parentheses sampled within the decoration. Another technical issue that arises due to the SMILES language is that sampled cycles identifiers ($1, 2, 3, ...$) should be different than the ones used in the scaffold, to ensure the respect of the given scaffold in corner cases where a cycle opened before an open position is closed after said open position.

\begin{algorithm}[H]
\SetAlgoLined
\KwResult{Smiles string with completed decoration}
 \KwIn{hidden state $h$}
  $h'_{0} = h$ \; 
  opened = 1 \;
  closed = 0\;
  $t = 1$\;
 \While{opened $>$ closed}{
  Sample $x_{t}$ from $P(x\mid h'_{t-1})$ and update $h'_{t-1}$ to $h'_{t}$ \;
  \uIf{x = '('}{
    opened += 1 \;
  }
  \uElseIf{x = ')'}{
    closed += 1 \;
  }
  $t = t+1$ \;

 }
 \caption{Decorating a scaffold on a given branching position}
\end{algorithm}

Additional refinements are possible, such as specifying the beginning of the branch (e.g “(CN*)” instead of “(*)”), or restraining the branch to be a linear fragment (by forbidding the opening of new branches and cycles).

\begin{figure}[H]
    \centering
    \includegraphics[scale=0.5]{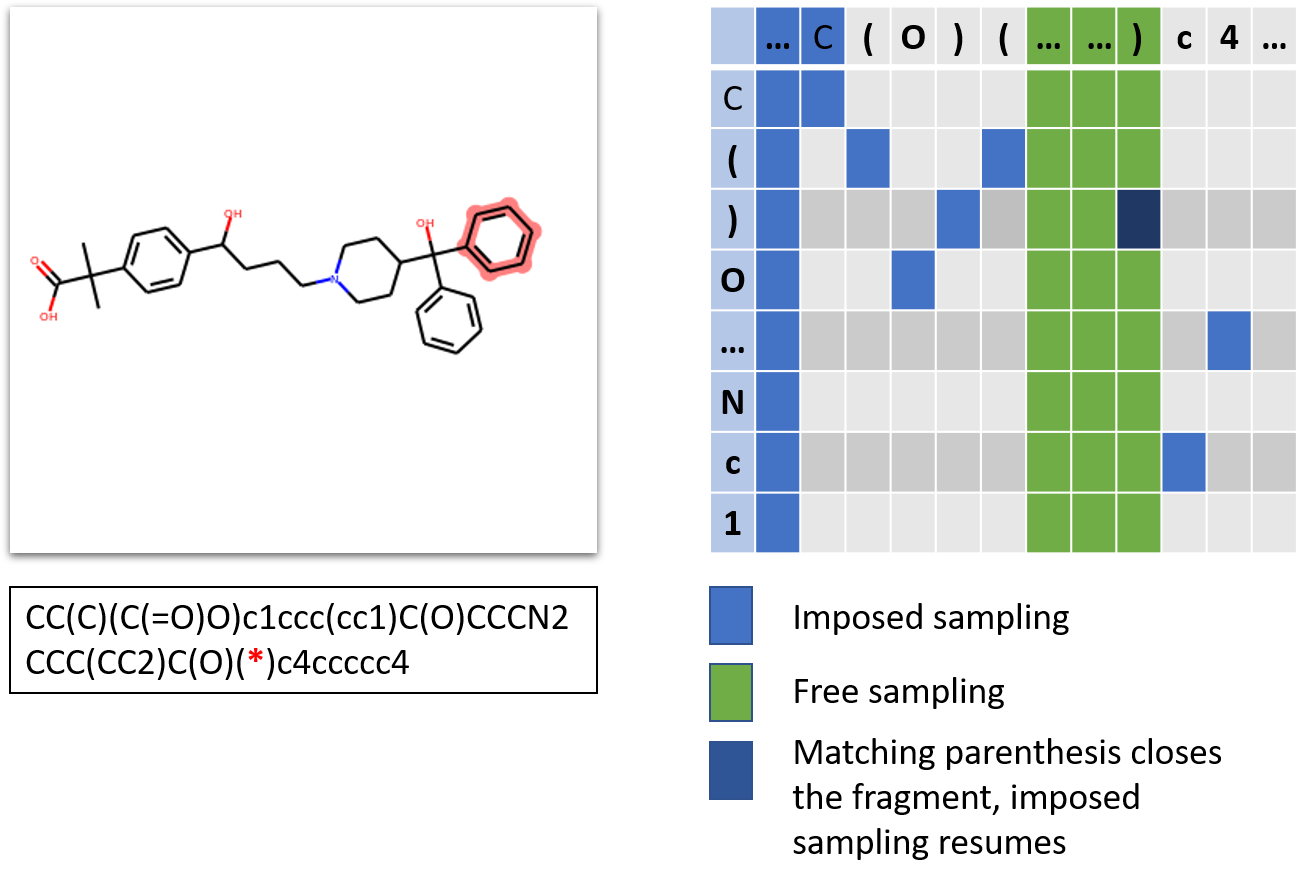}
    \caption{Sampling a new branch}
    \label{fig:Sampling branch}
\end{figure}

If branched decorations represent the majority of modifications allowed on a scaffold during lead optimization phases, it is sometimes interesting to allow more profound changes on the scaffold by performing scaffold hopping \cite{Bhm2004}.
To tackle this particular task, we investigate the possibility to have an open position within a linker between cycles.
This problem presents different difficulties. First, contrary to branched decorations, there is no clear indicator for stopping sampling and resuming reading the pattern.
Thus, the stopping criterion will necessarily be arbitrary. We implement it under the form of a user-defined probability distribution on the length of the added fragment.
Furthermore, as the end of sampling is arbitrarily decided rather than chosen by the RNN, we need to keep track of opening and closing parentheses as well as cycles to ensure that branches and cycles are completed within the added fragment before stopping sampling.
The stopping criterion for sampling is therefore a combination of a specified probability-distribution and if the sampled fragment doesn’t contains uncompleted cycles and branches. We do not look into the task of modifying existing cycles, which is more complex.

\begin{algorithm}
\SetAlgoLined
\KwIn{hidden state $h$, distribution on linker size $P_{size}$} 
\KwResult{Smiles string with completed linker}
 
 $h'_{0} = h$ \;
 Sample $n_{char}$ from $P_{size}$ \;
 opened = 0 \;
 closed = 0 \;
 step = 0 \;
 cycle = False \;
 \While{\textrm{cycle} \textbf{or} \textrm{step} $< n_{char}$ \textbf{or} \textrm{opened} $>$ \textrm{closed}}{
  Sample $x_{t}$ from $P(x|h_{t})$ and update $h'_{t-1}$ to $h'_{t}$\;
  \uIf{$x_{t}$ = '('}{
   opened = opened + 1 \;
  }
  \uElseIf{$x_{t}$ = ')'}{
   closed = closed + 1 \;
  }
  \uElseIf{$x_{t} \in \{1, 2,\dots, 9\}$}{
      \uIf{\textrm{corresponding cycle not opened}}{
       cycle = True \;
       keep track of opened cycle \;
      }
      \uElse{
       close corresponding cycle \;
       }
       }
       \uIf{\textrm{no cycle still opened}}{
       cycle = False \;
      }
  
 step = step + 1\;
 }

 \caption{Scaffold hopping by linker completion}
\end{algorithm}

\begin{figure}[H]
    \centering
    \includegraphics[scale=0.5]{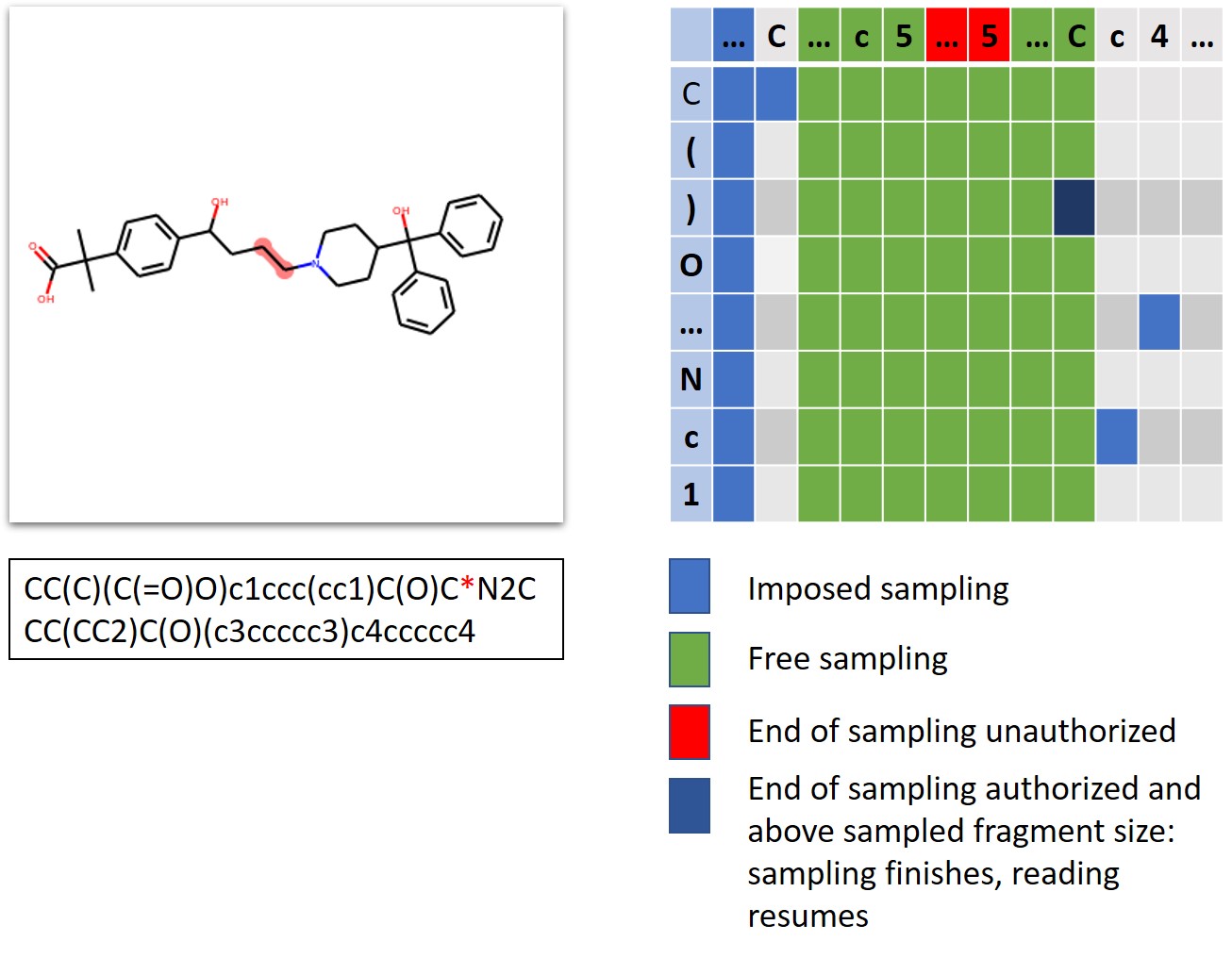}
    \caption{Scaffold hopping by sampling a new linker}
    \label{fig:Sampling linker}
\end{figure}

The ability to modify the core of a molecule allows our method to tackle more than simply scaffold decoration and thus to perform scaffold-hopping as well (see figure S1 for sampled molecules with a modified core, starting from the scaffold in Figure \ref{fig:Sampling linker}).

Finally, a simple use case nonetheless of practical interest is when open positions are actually restricted to a small number of discrete choices. During lead optimization campaigns, small variations on the scaffold are often encountered. Hence, we would like to keep this variability when designing molecules, while restraining our choices to the few possibilities that exist in this given context.
The SMARTS \cite{smiles} language, an extension of SMILES, deals with discrete choices with the following syntax: $[x_{1},x_{2},...,x_{k}]$ represents a discrete choice between SMILES characters $x_{1},x_{2},...,x_{k}$. We use this syntax for sampling between a finite set of discrete choices.
The sampling procedure is straightforward: restrict the possible tokens to those present in the discrete choices, renormalize the probability distribution and then sample.
Instead of drawing the next token from $P(x|h_{t})$, we sample it from: 
\begin{equation}
    Q(x) = \textrm{softmax}[P(x|h_{t}) \cdot \sum_{c \in \textrm{choices}}^{} \delta x,c]
\end{equation}

By design, our method can deal with multiple open positions of different nature within the same scaffold. This is important as scaffold constrained lead optimization often deals with multiple open positions at once.

\section{Experiments}
The objective of the following experiments is to verify the ability of our method to perform the major tasks that can be encountered in the context of scaffold-constrained molecular generation. We first check the ability of our method to perform scaffold-constrained generation on previously unseen scaffolds. We then assess the capacity of our method to generate, if needed, analogs to a given chemical series, in a focused learning task. Finally, we measure its performance in optimization of molecules with given constraints by benchmarking it in different scaffold-constrained goal-oriented scenarios. 

We rely on different sets of molecule for training and validation of the method. 
Following Olivecrona et al. \cite{article} 1.2 million molecules extracted from the ChEMBL database and filtered by drug-likeness related criteria, such as size, atoms, absence of macrocycles are used as a training set. 
To validate the ability to generate molecules around new scaffolds, we look into the SureChEMBL \cite{SureChemBL} database, a database of patented molecules. Clustering molecules by their Bemis-Murcko scaffold \cite{Bemis1996}, we extract 18 chemical series with 18 (see figure S2) different scaffolds.
Those scaffolds are chosen for validation as they 
\begin{itemize}
    \item are sampled from real life drug discovery projects 
    \item where not present in the training set (we remove any molecule that has one of the 18 scaffolds from the training set)
\end{itemize}

To study the ability to explore a focused region of chemical space and design analogs to a given series, we also isolate the largest (93 molecules) among the extracted chemical series.
This yields 17 scaffolds for the validation set, and one scaffold for the focused learning validation set.
Below, we refer to the molecules extracted from ChEMBL as the training set, the chemical series from SureChEMBL as the validation set, and the chemical series referenced above as the focused learning validation set. 

To implement our model, we build on the existing codebase released by \citeauthor{article} that already includes a SMILES based RNN, and with which many researchers are already familiar.
The RNN used in the subsequent tasks are either trained on the training set (onwards named “Generic RNN”) or on the focused learning validation set chemical series (named “Focused RNN”).

The rationale for using the Generic RNN in most applications is that the training set is comprised of diverse drug-like compounds, and that an RNN trained on it should be able to explore a large and varied chemical space. For the focused learning task, the goal is to assess the ability to generate close analogs to a given chemical series, and therefore the Focused RNN trained specifically on this chemical series is used.

For the task of generating molecules around an unseen scaffold, we use the Generic RNN to design molecules conditioned on the different scaffolds in the validation set. Major metrics used in benchmarking suites \cite{Brown2019,DBLP:journals/corr/abs-1811-12823} for \textit{in-silico} molecular generation are evaluated. The main goal is to ensure the ability to generate valid and unique SMILES, as well as to assess whether physico-chemical properties are similar to those of drug-like compounds.  
For the focused learning task, we compare distributions of molecules generated by the Generic RNN and the Focused RNN with the training set, validation set and the focused learning validation set.
As for goal-oriented benchmarks, we begin with the DRD2 target to provide a fair comparison with prior work \cite{ArsPous2020} \cite{Li2019} on scaffold constrained generation and investigate the ability of our method to generate predicted DRD2 actives on the different validation scaffolds studied in \cite{ArsPous2020}. To generate optimized molecules, we rely on a state-of-the-art Reinforcement Learning procedure, hill-climbing \cite{Brown2019}.
We then benchmark our method with the MMP-12 series \cite{Pickett2010} which is a large publicly available industrial lead-optimization dataset. Scaffold constraints are present in the dataset and explicitly mentioned in the original work that released the data, which supports our statement that they are common within drug-discovery lead optimization campaigns. After building a QSAR model on $\textrm{pIC}_{50}$ values, we compare our model against SMILES based LSTM with hill-climbing \cite{Brown2019},  a state-of-the-art \textit{in-silico} \textit{de-novo} design algorithm, in the task of generating predicted actives with the required scaffold.

\subsection*{Generating molecules with new scaffolds}

To verify the ability of our method to generate novel molecules around unseen scaffolds, we perform classic distribution learning benchmarks on sets of molecules designed around the scaffolds from the validation set. 
For each of the 17 scaffolds, we generate 10 000 SMILES. 
We first compute the proportion of (i) valid and (ii) unique SMILES. Validity is defined as whether the SMILES defines a valid molecular structure and is checked with the RDKit \cite{rdkit}. We shall note that validity ensures that the SMILES is syntactically valid (all rings and branches closed, no illegal atom types) but does not mean that the molecule will be necessarily synthetizable in a wet-lab. We also compute several physico-chemical properties for each valid molecule generated: calculated logarithm of partition coefficient(logP), Molecular Weight (MW), Synthetic Accessibility Score \cite{Ertl2009} (SAS), Quantitative Estimate of Drug-Likeness \cite{Bickerton2012} (QED), numbers of H donors (HBD) and acceptors (HBA). All properties were computed with the RDKit \cite{rdkit}. We group molecules generated for each of the different scaffolds together and compare the distributions of those properties between generated molecules, the training set and the validation set.

\begin{figure}[H]
    \centering
    \scalebox{.45}{\input{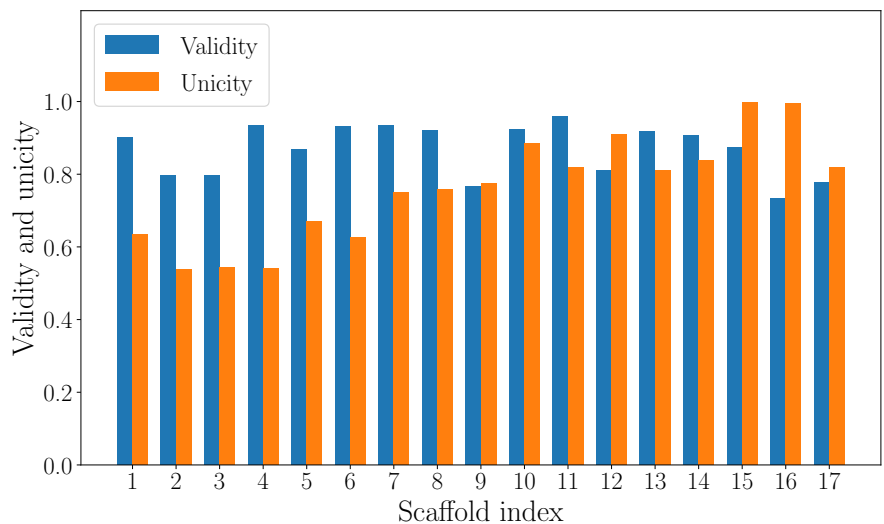}}

    \caption{Proportions of valid and unique molecules out of 10000 generated for each validation scaffold}
    \label{fig:val_unique}
\end{figure}

The proportion of unique and valid molecules for each scaffold in the validation set are given Figure \ref{fig:val_unique}, where the scaffolds are ordered by increasing number of open positions. Validity and uniqueness proportions (calculated over 10 000 SMILES for each scaffold), shown in Figure \ref{fig:val_unique}, are on par with the best scores obtained by various generative models \cite{Brown2019}.
Scaffolds in the validation set are ordered by ascending number of open positions; uniqueness proportion increases with the number of open positions, which matches the intuition that the more possibilities there is to modify the scaffolds, the more diverse the generated SMILES will be for a given scaffold.

\begin{figure}[H]
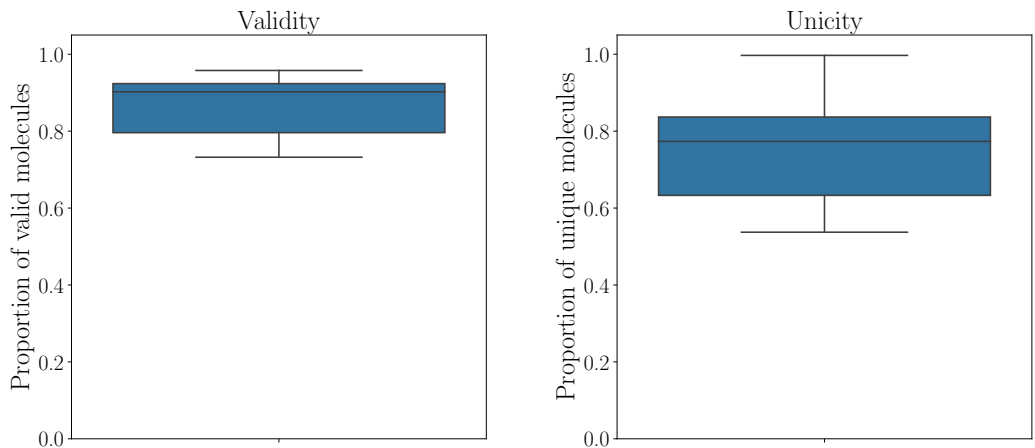

    \centering
    \scalebox{.4}{\input{figures_experiments/validity.pgf}}
    \scalebox{.4}{\input{figures_experiments/unicity.pgf}} 
    \caption{Validity and uniqueness proportions (min, lower quartile, median, upper quartile and max) across the 17 validation scaffolds}
    \label{fig:validity_unicity}
\end{figure}

We then perform comparison of distributions of various properties as a sanity check to ensure that generated molecules are similar to the drug-like molecules of the training and validation set.

\begin{figure}[H]
    \centering
    \scalebox{.65}{\input{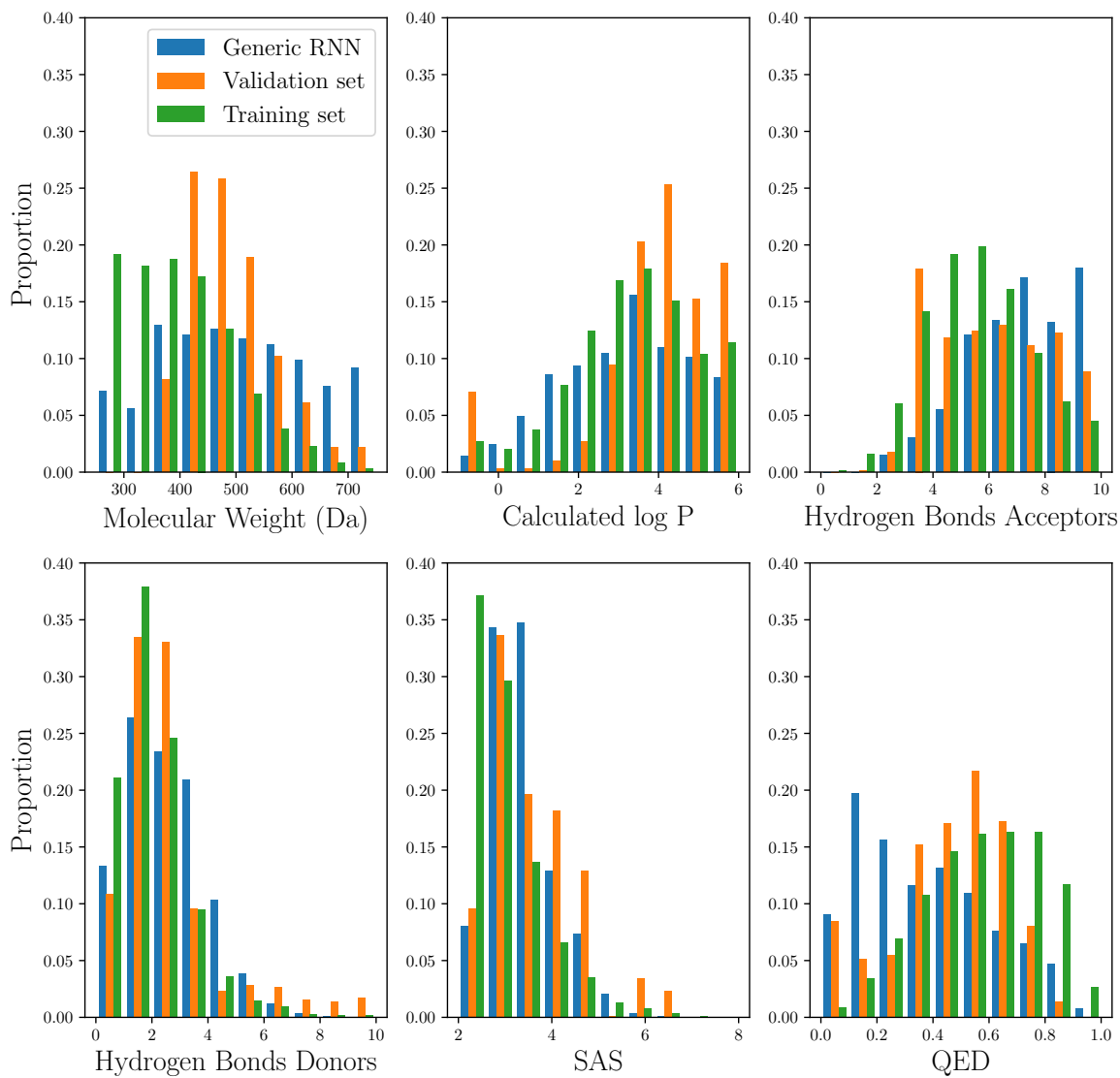}}
    \caption{Histogram of properties across generated molecules, training and validation set. For molecular weight and ClogP, values outside the [250,750] g.mol$^{-1}$ and [-1,6] ranges are not shown, and bars are the extremities accumulate values outside those ranges.}
    \label{fig:Properties}
\end{figure}

We note that properties are distributed similarly for the different sets of molecules, suggesting that generated molecules populates a similar property space as the training and validation set. An interesting fact is that generated molecules have lower QED scores; this is rather intuitive as molecules from training and validation set are actual molecules implying the bias that they were necessarily synthesizable and considered interesting enough in a drug discovery context so that their synthesis was actually performed.

\subsection{Focused learning on a chemical series}

In the context of lead-optimization, we might like to narrow the search for optimized molecules in a focused chemical space. Therefore, we investigate whether we could generate close analogs to an existing chemical series. We use the Focused RNN, trained specifically on the focused learning validation set (comprised of a single chemical series), to generate molecules with the scaffold of the focused learning validation set. 
For comparison, we also generate molecules with the same scaffold but using the Generic RNN. 

\begin{figure}[H]
    \centering
    \scalebox{.55}{\input{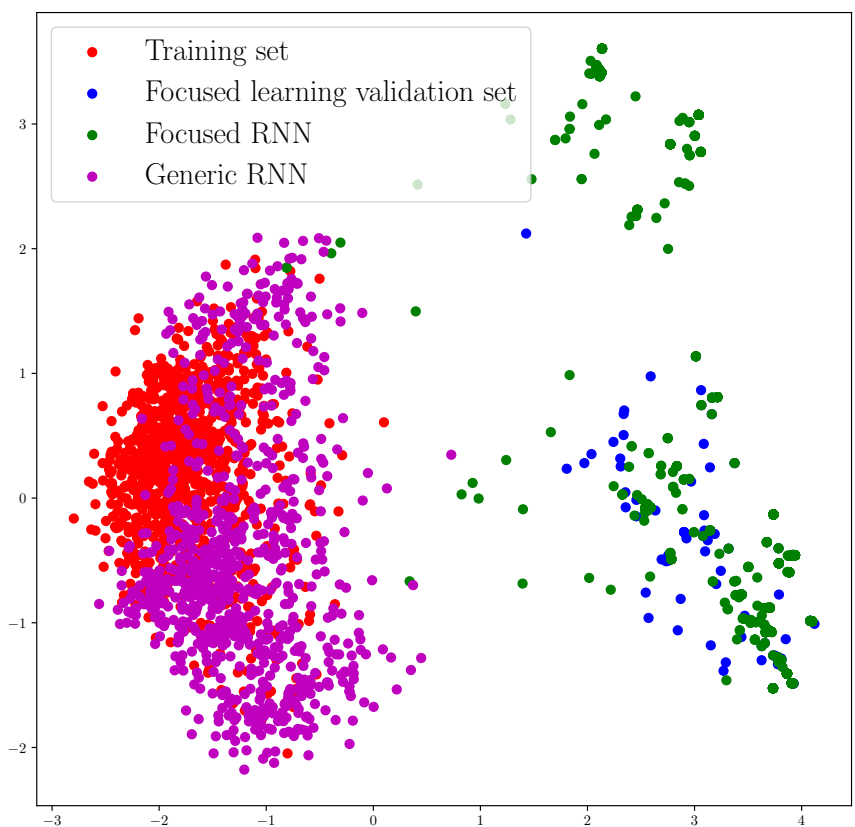}}
    \caption{PCA of generated molecules with Focused and Generic RNN compared with training set and focused learning validation set}
    \label{fig:PCA}
\end{figure}

We then perform dimensionality reduction to 2D with PCA on ECFP4 fingerprints.
Generated molecules are plotted in Figure \ref{fig:PCA} against the focused learning validation set and the training set.

Molecules generated with the Generic RNN overlap almost completely with the training set, while molecules generated with the Focused RNN overlap with the Focused learning validation series. This indicates that using a focused RNN allows to sample a chemical space close to a chemical series of interest.

It should be noted that while dimensionality reduction such as PCA on fingerprints is commonly used to compare distributions of molecules \cite{doi:10.1021/acscentsci.7b00512}, the question of whether this is the most relevant approach (especially as fingerprints are high-dimensional and binary) is still open, and should be kept in mind, as well as the fact that we are also comparing molecules with a shared substructure.

\subsection{Generating DRD2 actives}

Benchmarking our model on goal-oriented tasks is the main focus of our experiments. A benchmark for goal-oriented scaffold-constrained generation was proposed by \citeauthor{ArsPous2020}.
We therefore start by tackling the same task, which is generating predicted actives on the DRD2 target with specific scaffolds. We use the same 5 validation scaffolds. First, we assess unicity and validity proportions of generated molecules to ensure that our method generalizes well to those scaffolds. 
Then, for each scaffold, we run a reinforcement learning procedure, hill-climbing, with the objective of generating predicted actives.
Activity prediction is done with the QSAR model from \cite{article}. After ten epochs dedicated to learning to generate predicted actives, the best 50 molecules are kept.

\begin{table}
  [H] 
  \begin{tabular}
      {|p{5cm}|p{3cm}|p{3cm}|p{3cm}|} \hline \textbf{Scaffold} & \textbf{Valid molecules} &  \textbf{Unique molecules} & \textbf{Predicted active molecules out of 50 best} \\[5ex]
      \hline 
      \shiftright{0pt}{\raisebox{-1.3\height}{
      \includegraphics[scale=0.27]{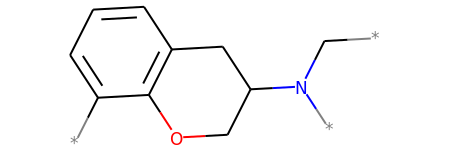}}} & \begin{center}86\%\end{center} & \begin{center}84\%\end{center} & \begin{center}\textbf{100\%}\end{center} \\[10ex]
      \hline
      \shiftright{6pt}{\raisebox{-1.7\height}{
      \includegraphics[scale=0.3, left]{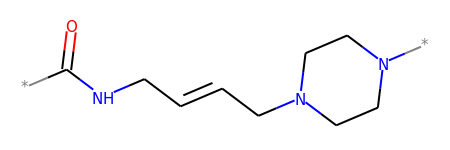}}} & \begin{center}82\%\end{center} & \begin{center}92\%\end{center} & \begin{center}\textbf{100\%}\end{center}  \\[20ex]
      \hline
      \shiftleft{5pt}{\raisebox{-1.0\height}{
      \includegraphics[scale=0.33,left]{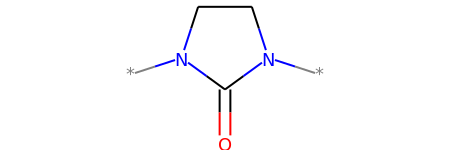}}} & \begin{center}86\%\end{center} & \begin{center}96\%\end{center} & \begin{center}\textbf{100\%}\end{center}  \\[8ex]
      \hline
      \shiftleft{7pt}{\raisebox{-0.9\height}{
      \includegraphics[scale=0.4,left]{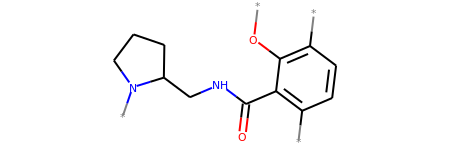}}} & \begin{center}92\%\end{center} & \begin{center}90\%\end{center} & \begin{center}\textbf{100\%}\end{center} \\[10ex]
      \hline
      \raisebox{-0.9\height}{
      \includegraphics[scale=0.3,left]{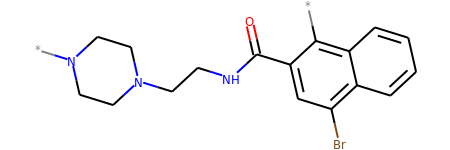}} & \begin{center}98\%\end{center} & \begin{center}32\%\end{center} & \begin{center}\textbf{100\%}\end{center} \\[8ex]
      \hline
  \end{tabular}
  \caption{Metrics on distribution learning and goal-oriented learning for DRD2 validation scaffolds}
   \label{table:DRD2}
\end{table}

Comparison with the Reinvent Scaffold Generator was not direct \footnote{comparison with \cite{Li2019} is not provided as we failed to make use of the code released to replicate the findings of this work}, as no insight was given of the total number of molecules generated for each scaffold. As mentioned previously, we made the choice to assess the top 50 best molecules generated. Nonetheless, we find that our model is able to generate a much higher proportion of predicted actives for each scaffold. This is achieved even without access to known actives on DRD2 with different scaffolds, unlike in \citeauthor{ArsPous2020} where training with actives is required.
Furthermore, relying on a much simpler architecture translates into a much higher throughput for generated molecules (roughly 100 times faster in CPU-time). Contrary to the Reinvent Scaffold Generator, we also do not require the use of a special preprocessing algorithm or of specific pretraining. On the other hand, we require a Reinforcement Learning procedure to be ran, though it comes at a rather cheap  computational cost (in the limit that the property of interest can be computed efficiently). This allows us to be much more efficient in finding molecules optimizing a given objective.

\begin{table}
    \begin{tabular}{|p{3cm}|p{6cm}|p{6cm}|}
    \hline
        & \begin{center}CPU time for generating 1000 molecules\end{center} & \begin{center}Molecules generated per second (CPU time)\end{center} \\
    \hline
    \begin{center}Reinvent Scaffold Decorator\end{center} & 
    \begin{center}620.9 seconds\end{center} & \begin{center}1.6 molecules/seconds\end{center} \\
    \hline
     \begin{center}\textbf{Scaffold-constrained generator}\end{center} & \begin{center}\textbf{7.02 seconds}\end{center} & \begin{center}\textbf{143 molecules/seconds}\end{center} \\
    \hline
    \end{tabular}
     \caption{Speed of generation comparison}
   \label{table:CPU}
\end{table}
\subsection{A lead-optimization use case: the MMP-12 series}

Experiments on the DRD2 target shows that our method can design predicted actives on a biological target while satisfying scaffold constraints. To provide further comparison, we designed a novel benchmark on the MMP-12 series, a publicly available industrial lead-optimization dataset, and assessed how scaffold-constrained generation fares against a state-of-the-art generative model. As goal-oriented models primary target is lead-optimization, designing a benchmark that resembles the industrial problem we’d like to solve makes sense. 
We compare our method and classic SMILES-based RNN with hill climbing reinforcement learning \cite{Brown2019}. For each method, 10 runs of reinforcement learning are launched, with the exact same optimization procedure and number of steps to ensure a fair comparison. For each run, the top 50 molecules are kept. For each molecule, given its predicted  $\textrm{pIC}_{50}$, the score is computed as:

\[
\begin{cases}
    \textrm{max}(1, 1 - (7.5-\textrm{pIC}_{50})/7.5),& \text{if scaffold constraints are met} \\
    0,              & \text{otherwise}
\end{cases}
\]

Molecules that have predicted $\textrm{pIC}_{50}>7.5$  while respecting the scaffold constraint have therefore a score of 1.
 
We report for each method,  the percentage of actives (predicted $\textrm{pIC}_{50}>7.5$) with substructure (therefore matching 2/2 of the project’s requirements), only actives (1/2 requirements) , only matching substructures (1/2 requirements as well) and none of the two (0/2 requirements met).

\begin{table}
    \begin{tabular}{|p{3cm}|p{3cm}|p{3cm}|p{3cm}|p{3cm}|}
  \hline
   & \begin{center}\textbf{Right scaffold, active (2/2)}\end{center} & \begin{center}Right scaffold, not active (1/2)\end{center} & \begin{center}Active, without scaffold (1/2)\end{center} & \begin{center}Not active, without scaffold (0/2)\end{center} \\
  \hline
  \begin{center}Classic RNN\end{center} & \begin{center}\textbf{0\%}\end{center} & \begin{center}56\%\end{center} & \begin{center}42\%\end{center} & \begin{center}2\%\end{center} \\
  \hline
  \begin{center}Scaffold-constrained generator\end{center} & \begin{center}\textbf{23\%}\end{center} & \begin{center}77\%\end{center} & \begin{center}0\%\end{center} & \begin{center}0\%\end{center} \\
  \hline
    \end{tabular}
    \caption{Generating actives on the MMP-12 series}
    \label{table:MMP-12}
\end{table}
With our method, we find 23\% of molecules that satisfy 2/2 requirements for the project while no satisfying molecules are found with classic SMILES-based RNN. This discrepancy is probably due to the fact that the chemical space where the structure constraint is met is a very sparse subspace of chemical space, which hinders the performance of reinforcement learning. This leads the classic RNN to struggle generating molecules within this chemical space while optimizing activity. On the contrary, our method generates by design only molecules that meet the scaffold requirement, allowing the optimization procedure to focus only on the true objective. 
Amongst the molecules discovered by our method are present some experimentally validated actives that were part of the held-out validation set, yielding yet another confirmation that inverse QSAR powered by generative models can discover experimentally validated molecules.

\begin{figure}[H]
    \centering
    \begin{center}
    \includegraphics[scale=0.5]{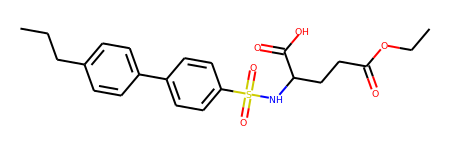}
    \end{center}
    \caption{Re-discovered active from the held-out validation set, with the scaffold constraint highlighted}
    \label{fig:MMP_12_active}
\end{figure}

This experiment shows that, by releasing the structure constraint (as it is built in our model), our method can search much more efficiently for optimized molecules (w.r.t predicted activity) within a subspace of molecules with a required scaffold.
This experiment also shows that optimization under structure constraints seems to be a difficult task for generative models. For instance, on the Guacamol \cite{Brown2019} benchmarking suite, optimization on multiple objectives is not problematic and high scores are achieved on a wide range of tasks by SMILES-based RNN. Yet, having only two objectives in this task with one being a scaffold constraint, classic SMILES-based RNN struggles to generate optimized molecules, which gives a significant advantage to  scaffold constrained generation. As scaffold constraints are common within drug discovery projects, we think that our method could therefore prove to be very useful in this context.

\section{Conclusion}

Applying generative models to drug-discovery can be used in lead optimization tasks, that often require to respect scaffold constraints. Those constraints are mainly present for preserving biological activity and staying in known SAR domains for the optimized properties, as well as for synthesizability and optimization of the synthesis process. Including those constraints in the generative process of a model is a difficult problem, and that has potential practical impact on the applications of generative models to drug-discovery. 
In this work, we investigate how a well-known model, SMILES-based RNN, can be slightly modified to achieve different scaffold-constrained generative tasks. This is possible thanks to a modified sampling procedure. Our approach for scaffold constrained generation thus doesn’t require designing a new model, or even retraining it. Furthermore, all previous works on reinforcement learning for molecular properties optimization with SMILES-based RNN stay applicable. 
Using distribution learning benchmarks, we show that our method can generalize across unseen scaffolds, and can also generate molecules in a focused fashion.
The validation scaffolds used for this task are extracted from SureChEMBL and thus derived from real lead optimization chemical series. On scaffold constrained goal-oriented benchmarks, our method largely outperforms state-of-the-art \textit{de-novo} design algorithms. Our approach was able to generate new predicted actives on the DRD2 target, without specific pretraining. Furthermore, we showed that it outperformed classic SMILES based reinforcement learning for designing predicted actives on the MMP-12 series, proposing held-out experimentally validated actives. 
We also show that by design, reinforcement learning methods are applicable in this context. This enables our method to use state-of-the-art algorithms for goal-oriented tasks, and we show strong performance on scaffold constrained in-silico molecular optimization.
Our model also goes beyond simple scaffold decoration. It can provide low level control on the way the scaffold is completed, and can also be used in other tasks such as scaffold hopping.
Limitations of the method include the need to handcraft the scaffold constraints in SMILES format, as well as the fact that in particular instances, specificities of the SMILES language requires manual overriding of sampled cycles to ensure the respect of scaffold constraints.
Overall, we believe our method shows a real practical interest for scaffold constrained optimization tasks that make the most of actual lead optimization challenges in drug discovery. Coupled with the fact that we rely on a well-known and already widely adopted model, we expect that it will benefit researchers looking to apply generative models for lead optimization tasks.



        
\newpage
\bibliography{achemso-demo}
\begin{figure}[H]
    \centering
    \begin{center}
    \includegraphics[width=8.25cm,height=4.45cm,keepaspectratio]{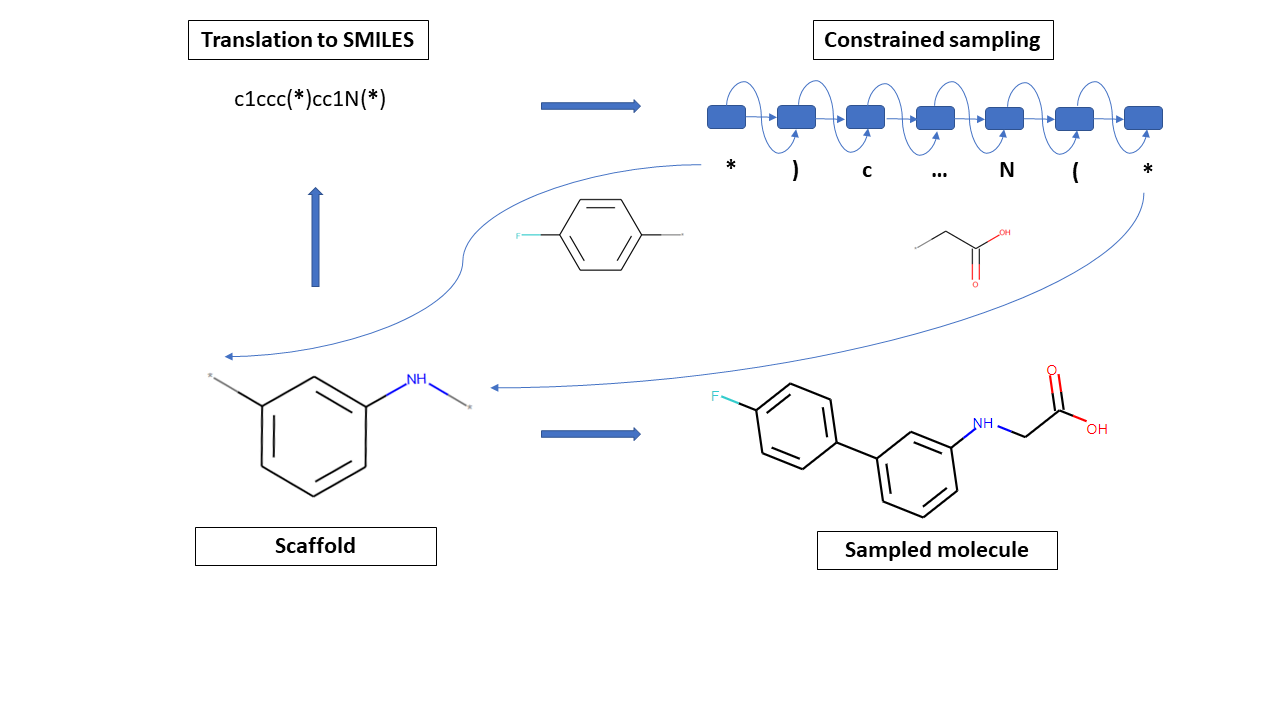}
    \end{center}
    \caption{For Table of Contents Only}
    \label{fig:TOC}
\end{figure}
\end{document}


\section{Data curation and software availability}

\subsection{Code availability}

All software and data to reproduce the results of this paper are available at: 

\url{https://github.com/maxime-langevin/scaffold-constrained-generation}.

To implement the methods presented above, we build on the existing codebase of Olivecrona and al.
, available at \url{https://github.com/MarcusOlivecrona/REINVENT}.
The different experiments are reproduced in Jupyter notebooks
available in our codebase. An extra notebook showing basic usage for researchers interested in simply using our method without necessarily reproducing our results is also available.
\subsection{ChEMBL dataset}

The ChEMBL$^{26}$
database is often used to train generative models of drug-like molecules.
To train our RNN, we use a preprocessed version of the ChEMBL$^{13}$ 
where only molecules having between 10 and 50 heavy atoms and comprised of elements $\in \{H, B, C, N, O, F, Si, P, S, Cl, Br, I \}$ were kept. The original filtered ChEMBL dataset can be found and downloaded at: \url{https://github.com/MarcusOlivecrona/REINVENT/blob/master/data/ChEMBL_filtered}, and in our repository at \url{https://github.com/maxime-langevin/scaffold-constrained-generation/data/ChEMBL_filtered}. Furthermore, we filtered the dataset to exclude molecules having one the 17 validation scaffolds as a substructure, yielding the final dataset at \url{https://github.com/maxime-langevin/scaffold-constrained-generation/data/ChEMBL_without_sureChEMBL.smi}.

\subsection{SureChEMBL dataset}

The SureChEMBL$^{29}$ 
database is comprised of patented compounds. The database can be downloaded at
\url{https://chembl.gitbook.io/chembl-interface-documentation/downloads}. 
34000 compounds were exctracted from SureChEMBL v2019.10.01. Compounds were clustered by Bemis-Murcko scaffold$^{30}$ 
and 18 chemical series (every molecule in each series having the same scaffold) were kept to be used as a validation set.
The molecules in the 18 series can be found at \url{https://github.com/maxime-langevin/scaffold-constrained-generation/data/SureChEMBL/200323_SureChemBL_dataset_636.sdf}, and the 18 scaffolds at \url{https://github.com/maxime-langevin/scaffold-constrained-generation/data/SureChEMBL/surechembl_scaffolds.sdf}.

\subsection{DRD2 dataset}

The full DRD2 dataset can be found at
\url{https://github.com/undeadpixel/reinvent-scaffold-decorator/blob/master/training_sets/drd2.excapedb.smi.gz} . The scaffolds used for the goal-directed benchmark are the ones used in Ar\'us-Pous and al.
, and can be found in our codebase at \url{https://github.com/maxime-langevin/scaffold-constrained-generation/data/DRD2/drd2_scaffolds.sdf}.

\subsection{MMP-12 dataset}

The MMP-12 dataset was downloaded from the supplementary materials of Pickett and al. 
The dataset can be found at \url{https://github.com/maxime-langevin/scaffold-constrained-generation/data/MMP12/mmp12.csv}.

\section{Implementation details}

\subsection{Computation time benchmarks}
Computation time benchmarks were run on an Amazon EC2 p2.xlarge, and assessed the runtime of generating molecules using one CPU.

\subsection{Distribution learning benchmarks}
To assess distribution learning benchmarks, $10 000$ molecules were generated for each scaffold.
\subsubsection*{Validity}
The validity score for a scaffold is the ratio of the number of valid molecules, as defined in the RDKit 
, out of all $10 000$ generated molecules.

\subsubsection*{Unicity}
Out of the generated valid molecules, the number of unique molecules is computed as the ratio of molecules with distinct canonical SMILES string.

\subsubsection*{Physico-chemical properties}

All physico-chemical properties where computed using the RDKit. Properties were computed on the valid molecules out of the $10 000$ generated for each scaffold, and then grouped together. The overall distributions were plotted against the distributions of both the training and the validation set, in order to check that there was no striking dissimilarity between them. 
\subsection{Predicting DRD2 activity}
One of the major point of the DRD2 activity was to benchmark our method against the Reinvent Scaffold Decorator$^{22}$.
Thus, it seems natural to use the same QSAR model. As we weren't able to find this QSAR model within the codebase reproducing the experiments of the article, we used a QSAR model$^{13}$ 
used in a work from the same group on the DRD2 dataset, and we assumed that the QSAR model used in the two works was the same.
In our codebase, the model used for DRD2 activity prediction can be found at \url{https://github.com/maxime-langevin/scaffold-constrained-generation/data/clf.pkl}

\subsection{Predicting MMP-12 activity}
To predict activity on the MMP-12 target, the dataset was split into a training and a test set. Then, a random forest regression algorithm (implemented with Scikit-learn)  
was fitted on the training set with continuous targets (corresponding to the experimental $pIC_{50}$), and evaluated on the testing set. 
The evaluation yielded a coefficient of determination $r^2 = 0.84$. The QSAR model is accessible at \url{https://github.com/maxime-langevin/scaffold-constrained-generation/data/MMP12/final_activity_model.pkl}, and evaluation on the test set at \url{https://github.com/maxime-langevin/scaffold-constrained-generation/MMP12_experiments.ipynb}.

\subsection{Hill climbing procedure}
To optimize molecules in goal-oriented benchmarks, a hill-climbing procedure$^{4}$ 
was used, as it was shown to be overall the best method amongst different generative models. The algorithm can be summarized as a repetition of the following steps:
\begin{itemize}
    \item Generate 500 molecules
    \item Score them and keep the top 50 unique molecules
    \item Perform 10 rounds of log-likelihood maximization with the 50 best molecules 
\end{itemize}
Those steps are repeated 10 times in a row.
The code for performing hill-climbing can be found at \url{https://github.com/maxime-langevin/scaffold-constrained-generation/hill_climbing.py}.

\section{Figures}

\begin{figure}[H]
    \centering
    \includegraphics[scale=0.35]{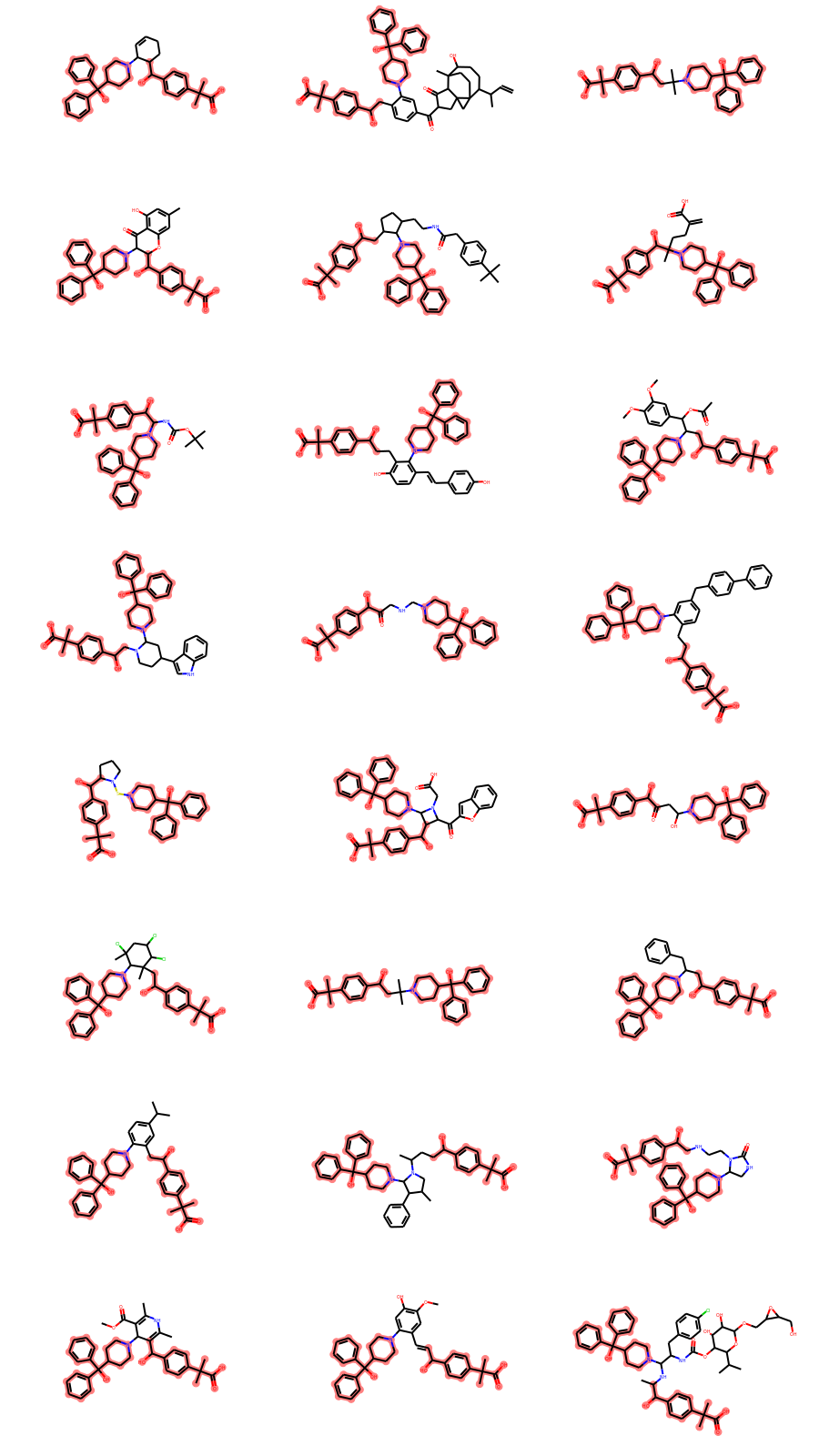}
    \caption{Molecules obtained after modifying the core of fexofenadine. The scaffold constraint is highlighted.}
    \label{supp:scaffold_hopping}
\end{figure}

\begin{figure}[H]
    \centering
    \includegraphics[scale=0.4]{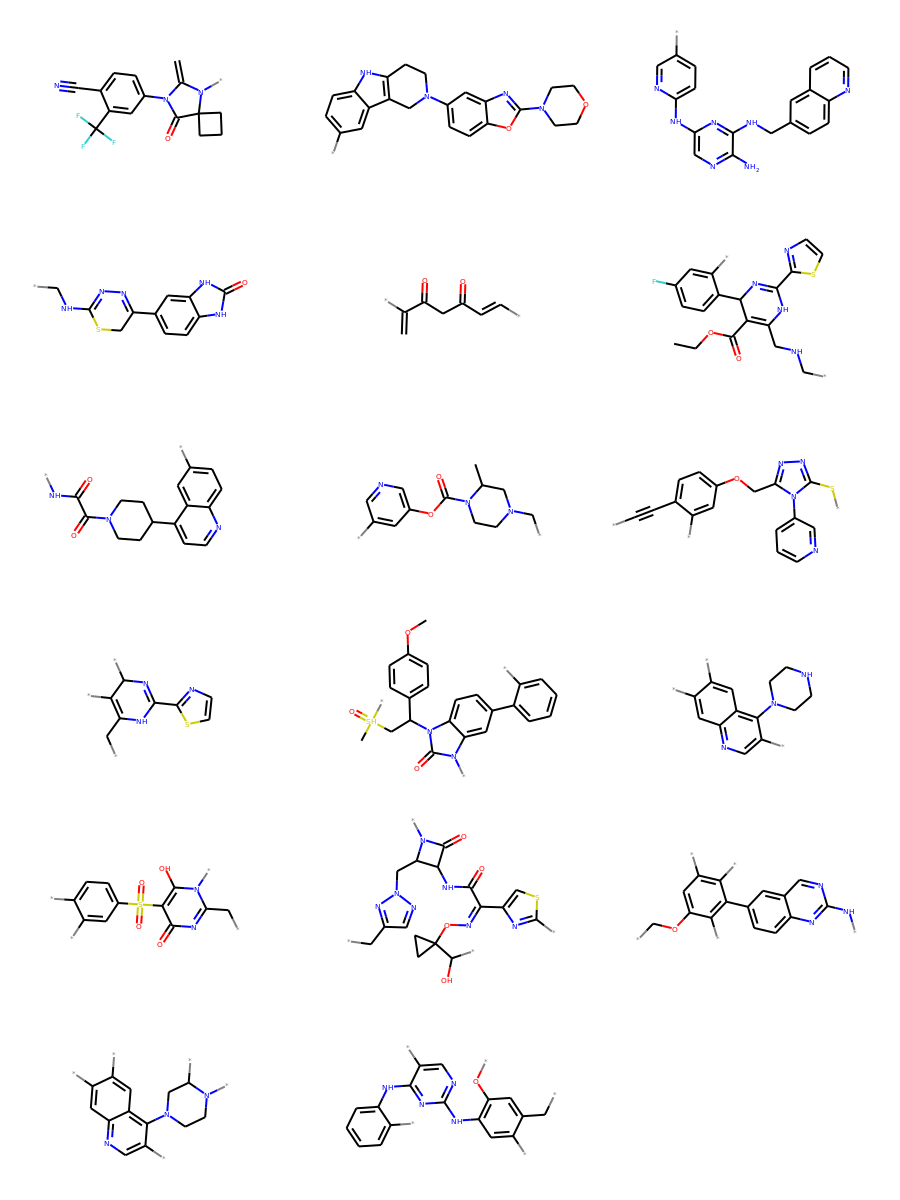}
    \caption{Validation scaffolds from SureChEMBL}
    \label{supp:scaffs_surechembl}
\end{figure}
